\documentclass[namedreferences]{solarphysics}
\usepackage[natbib]{spr-sola-addons} 
\usepackage{graphicx}
\usepackage{color}

\begin{document}

\begin{article}

\begin{opening}

\title{The Dynamic Spectrum of Interplanetary Scintillation: First Solar Wind Observations on LOFAR}

\author{R.A.~\surname{Fallows}$^{1,2}$\footnote{Moved to institute (1) from 1 February 2012}\sep
        A.~\surname{Asgekar}$^{1}$\sep
        M.M.~\surname{Bisi}$^{2}$\sep
        A.R.~\surname{Breen}$^{2}$\footnote{Deceased}\sep
        S.~\surname{ter-Veen}$^{3}$\sep
        on~\surname{behalf of the LOFAR Collaboration}\footnote{Full author list at http://www.astron.nl/authors-list-lofar-commissioning-papers}
        }
\institute{$^{1}$ ASTRON - the Netherlands Institute for Radio Astronomy, Postbus 2, 7990 AA Dwingeloo, The Netherlands \\
                 e-mail: fallows@astron.nl \\
                 e-mail: asgekar@astron.nl \\
           $^{2}$ Institute of Maths and Physics, Aberystwyth University, Aberystwyth, SY23 3BZ, Wales \\
                 e-mail: Mario.Bisi@aber.ac.uk \\ 
           $^{3}$ Department of Astrophysics/IMAPP, Radboud University Nijmegen, P.O. Box 9010, 6500 GL Nijmegen, The Netherlands \\
                 e-mail: S.TerVeen@astro.ru.nl \\
           }

\runningauthor{Fallows et al.}
\runningtitle{LOFAR Observations of the Solar Wind}

\begin{abstract}
The \textit{LOw Frequency ARray} (LOFAR) is a next-generation radio telescope which uses thousands of stationary dipoles to observe celestial phenomena. These dipoles are grouped in various `stations' which are centred on the Netherlands with additional `stations' across Europe. The telescope is designed to operate at
frequencies from 10 to 240\,MHz with very large fractional bandwidths (25-100\%). Several `beam-formed' observing modes are now operational and the system
is designed to output data with high time and frequency resolution, which are
highly configurable. This makes LOFAR eminently suited for dynamic spectrum
measurements with applications in solar and planetary physics. In this paper we
describe progress in developing automated data analysis routines to compute
dynamic spectra from LOFAR time-frequency data, including correction for the
antenna response across the radio frequency pass-band and mitigation of
terrestrial radio-frequency interference (RFI). We apply these data routines to
observations of interplanetary scintillation (IPS), commonly used to infer
solar wind velocity and density information, and present initial science
results.
\end{abstract}


\end{opening}


\section{Introduction}
\label{sec:intro}

The observation of interplanetary scintillation (IPS) -- the scintillation of compact radio sources due to density variations in the solar wind \citep{Hewishetal:1964} -- is an important tool for observing the solar wind.  Observations of IPS allow the solar wind speed to be inferred over all heliographic latitudes and a wide range of elongations from the Sun \citep[e.g.][]{DennisonHewish:1967}, giving a global perspective to point measurements from spacecraft.  The sensitivity of IPS to small, turbulent-scale, density variations also complements the larger-scale sensitivities of white-light observations from coronagraphs.

Advances in the study of IPS over the last decade or more allow such observations to be used to calculate three-dimensional (3D) sky maps of solar wind electron density and speed \citep[e.g.][]{Asaietal:1998, Jacksonetal:1998, Kojimaetal:1998}.  These maps and other detailed analyses of observations of IPS \citep[e.g.][]{Fallowsetal:2008a, Breenetal:2008, Bisietal:2007} are increasingly recognised as a valuable aid for tracking space weather events through the inner heliosphere \citep[e.g.][]{Jacksonetal:2010} and to the overall study of space weather prediction.

The \textit{LOw-Frequency ARray} (LOFAR -- summarised fully in Section \ref{sec:lofar}) is a major new-generation radio telescope operating in the 10--240\,MHz frequency range.  It consists of arrays of dipoles grouped into stations with a central `core' of stations in the Netherlands and, currently, eight international stations based in the UK, France, Germany and Sweden.  Although designed principally to be used as a single array, it is also possible to use the stations individually making it suitable for studies of IPS.  A particular advantage offered by LOFAR is the ability to observe with bandwidths of up to 48\,MHz with a high frequency resolution.  This capability is both necessary to identify and eliminate radio frequency interference (RFI) and useful to create dynamic spectra of IPS data, a tool that could provide new insights into solar wind microstructure and has not been available to most prior IPS observing instruments. 

This paper describes the principle science of IPS, details how LOFAR may be used as an instrument to observe it, the principle advantages LOFAR can offer as such an instrument and the progress made in obtaining observations of IPS. We discuss in particular on-going efforts to develop an automated Python-based software `pipeline' to produce relevant products from raw LOFAR data. Over time these tools
will be developed to enable the study of other similar phenomena, such as flare stars, planetary atmospheres and solar radio bursts.

The paper is laid out as follows:  The study of IPS is summarised in Section \ref{sec:ips}; Section \ref{sec:lofar} describes the LOFAR radio telescope and details how it may be used for IPS; the current state of a dynamic spectrum data pipeline developed for LOFAR observations is detailed in Section \ref{sec:rfi} and then some initial IPS results are presented in Section \ref{sec:results}.


\section{Interplanetary Scintillation (IPS)}
\label{sec:ips}

\begin{figure}
	\centering
	\includegraphics[width=10cm]{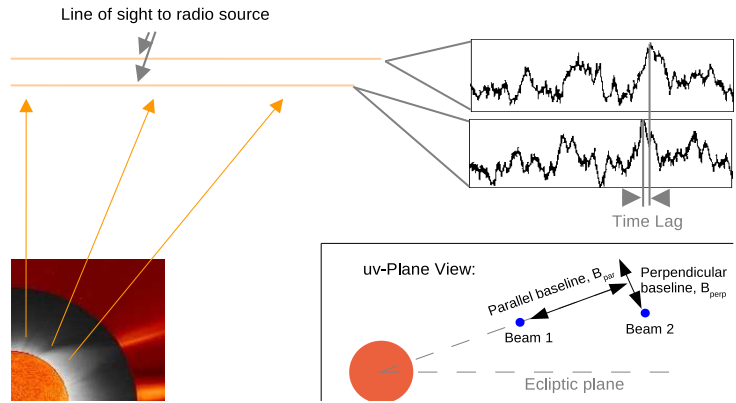}
\caption{Illustration of the geometry of the lines of sight from two telescopes
with respect to the Sun and the patterns of interplanetary scintillation
detected in each one. Main illustration is from above the ecliptic plane; inset
is in the uv plane (the projection of the plane on which the antennas lie to be perpendicular to the radio source direction) as viewed from the radio source.}
	\label{fig:ips}
\end{figure}

When two radio telescopes are used and their projected baseline on the
$u-v$ plane is close to the radial direction centred at the Sun, a high degree
of correlation may be observed between the scintillation patterns recorded at
the two telescopes \citep[e.g.][]{ArmstrongColes:1972b, Coles:1996}. The
time lag for maximum cross-correlation of the two simultaneously-taken radio
signals can be used to estimate the outflow speed of the density variations
producing the scintillation and, thus, a mean outflow speed for the solar wind
across the line of sight. Figure \ref{fig:ips} illustrates the geometry and the
scintillation patterns recorded by two telescopes.

\begin{figure}
	\centering
	\includegraphics[width=10cm]{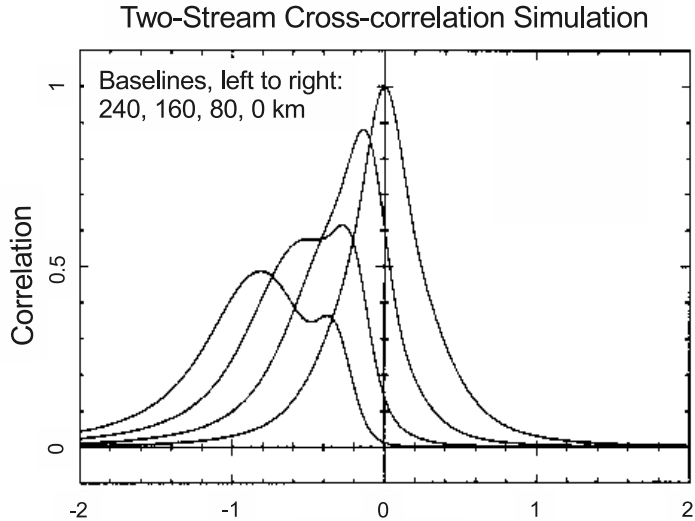}
\caption{Illustration of the effect on the cross-correlation function of
increasing the baseline from 0\,km to 240\,km between two radio telescopes when the lines of sight pass through both fast and slow solar wind streams.}
	\label{fig:ccfs}
\end{figure}

The lines of sight from the radio telescopes to the radio source may pass
through two or three solar wind streams each travelling at a different speed and having differing densities.
This may be observed directly if the baseline between the telescopes is
increased to several hundred kilometres. The cross-correlation function may then
display two or three distinct peaks, each at a time lag corresponding to individual
solar wind streams. The effect of increasing baseline length is shown in Figure
\ref{fig:ccfs}.  Here, four cross-correlation functions corresponding to
 baseline lengths of $0-240$\,km  are shown, for a model observation which assumes
the  presence of both fast and slow solar wind streams in the lines of sight.  The 
cross-correlation function at $0$\,km baseline is effectively an auto-correlation
 function of the two input signals. As the baseline is increased, the cross-correlation
function decreases in height and becomes first skewed before separating
into two distinct peaks corresponding to the two solar wind streams. As the
baseline is increased to very large distances, evolution
of the solar wind density structures will de-correlate the signals
significantly. However, cross-correlation is still evident on baselines of at
least 2000\,km \citep[e.g.][]{Breenetal:2006}.

More sophisticated analysis methods can be used to account for the line of sight
integration and provide more accurate estimates of solar wind speeds in the line of sight.  One method fits the results of a scattering model, which can assume up to three solar wind streams in the line of sight, to the observed auto- and cross- power spectra \citep{Fallowsetal:2008a, Bisietal:2007, Fallowsetal:2006, Coles:1996}; a second method uses a tomographic inversion of many observations taken over the course of a whole solar rotation to provide three-dimensional (3D) maps of solar wind speed and density via the use of a measure of the level of scintillation; \citep[e.g.][]{Asaietal:1998, Jacksonetal:1998, Kojimaetal:1998}.


\section{The Low Frequency Array (LOFAR)}
\label{sec:lofar}

LOFAR is designed and constructed by ASTRON, the Netherlands Institute for Radio Astronomy. It has facilities in several countries which are  collectively operated by the International LOFAR Telescope consortium.
LOFAR operates in the frequency range $10-240$\,MHz, offering a large collecting
area ($\sim 10^5$ m$^2$), and is comprised of thousands of dipole antennas
hierarchically arranged in stations which come in three different configurations
(Table \ref{tab:station_elements}). There are a total of 33 stations in the
Netherlands. These include a dense core of six stations, called the `Superterp',
near Exloo, and 18 `core' stations in the neighbourhood (baselines of  $\sim 2$\,km.
There are currently nine `remote' stations within the Netherlands, offering baselines
over $\sim 70$\,km, and eight international stations \mbox{up to $\sim 800$\,km}, with
the prospect of more to come. As will be discussed in more detail later, observations of IPS can utilise all of these stations to achieve a variety of diverse
goals. Details of system architecture and signal processing can be found in
\citet{deVosetal:2009} and a full description of LOFAR is in preparation by van Haarlem \textit{et al.}. 

LOFAR has two different types of antennas to cover the full frequency range. 
The low band antennas (LBAs) cover the frequency range $10-90$\,MHz,
although they are optimised for frequencies above 30 MHz. There are $48/96$
active LBA dipoles in each Dutch/international station (Table
\ref{tab:station_elements}). The high band antennas (HBAs) cover the
frequency range $110-240$\,MHz, and consist of 16 folded dipoles grouped into
tiles of $4 \times 4$ cross-dipoles each, which are phased together using an 
analogue beam-former within the tile itself.  It is possible to observe a radio 
source with a maximum bandwidth of $\sim 48$\,MHz.

With large fractional ($\sim100\%$) bandwidths, sophisticated multi-beaming
capabilities ($\sim100$ concurrent beams on the sky), and a large field of view,
LOFAR is a powerful instrument for surveys and routine monitoring of
variable sources. For more details the reader is referred to \citet{Stappersetal:2011},
where the capabilities of LOFAR for high-time-resolution beam-formed observations
are discussed in detail.

For  the study of IPS, LOFAR offers distinct advantages over other telescope systems:
\begin{itemize}
\item Multiple international stations spread around the central core offer a
greater number of useful cross-correlation observing opportunities at baseline
lengths which enable the full set of determinations (speeds of multiple solar wind streams for example) to be extracted.
\item The large bandwidth and excellent frequency resolution enable dynamic 
spectra to be calculated, which will undoubtedly provide a mine of additional
information on solar wind micro-structure.
\item With its enormous flexibility to observe many sources simultaneously, LOFAR offers a possibility of making detailed tomographic maps of solar wind speed and density.
\end{itemize} 

\begin{table}
\begin{tabular}{l c c c}
\hline \hline
Station Type  &  LBA (no.)  &  HBA tiles (no.) & Baseline (km) \\
\hline
Core          & 2$\times$48 & 2$\times$24      & $\le 2$       \\ 
Remote        & 2$\times$48 &  48              & $\ge 70$      \\
International &   96        &  96              & $\ge 300$     \\
\hline
\end{tabular}
\caption[h]{(After \citet{Stappersetal:2011}). Arrangement of elements in the
three types of LOFAR stations, along with their typical distance from the centre
of the array (baseline). In the Core and Remote stations there are 96 LBA
dipoles but only 48 can be beam-formed at any one time. For these stations, one
can select either the inner circle or the outer ring of 48 LBA
dipoles depending on the science requirements. The HBA sub-stations can be
correlated, or used in beam-forming, independently.}
\label{tab:station_elements}
\end{table}

\subsection{LOFAR configuration and various beam-formed modes of operation}

LOFAR offers a number of `beam-formed' modes, in which one can form single or
multiple beam  pointings in the direction of radio sources of interest using one or
more stations. Given the total bandwidth available for data transport, the total 
amount of observing bandwidth over all the beams from the stations is limited to
$48$\,MHz.  The various beam-formed modes probe timescales from seconds down to 
microseconds. Given that most LOFAR signal processing is carried out in software,
there are many ways in which the various parts of LOFAR (antennas, tiles, stations)
 can be combined to form beams.  The reader is referred to \citet{Stappersetal:2011}
 for a discussion on the different options for combining beam-formed data.

The term {\bfseries station beam} corresponds to the beam-formed by the sum of
all of the elements of a station. For any given observation there may be more
than one station beam and they can be pointed at any location within the wider
element beam. A {\bfseries tied-array beam} is formed by coherently combining
individual station beams (one for each station), which are looking in a particular
direction within each station beam. There may be more than one tied-array beam
for each station beam. Station beams can also be combined incoherently in order
to form {\bfseries incoherent array beams}.

The modes of relevance to IPS are:
\begin{itemize}
\item ``Fly's Eye'' (FE) mode, which allows individual station level beam-formed
data (complex voltages or coherent Stokes parameters) to be recorded separately rather than correlated before recording;
\item ``Tied Array'' (TA) beam mode, which forms single beams from multiple
stations (typically the Superterp).
\end{itemize}

The FE mode is necessary in order to study the cross-correlation between individual
stations as a result of IPS; it would not be possible to analyse the IPS time series' to obtain solar wind parameters if all stations used in a particular observation are correlated in the system as in the case for standard radio astronomy imaging.  The TA mode has been used in commissioning observations of IPS using the
Superterp. This offers an increased sensitivity and provides beam-width comparable 
to that from single large radio dishes, allowing a fair comparison with data from 
telescopes elsewhere.

The raw measurement for studying IPS is signal
intensity sampled at a high rate (most systems use a sampling rate of at least 50Hz); a sampling rate of 96Hz is used in the LOFAR observations presented here, chosen for system convenience.  In most
traditional radio telescope systems the signal intensity is integrated over the
bandwidth before being recorded. In the case of LOFAR, the dynamic spectrum is a
necessary intermediate step as well as being of scientific interest in itself.

The reason for recording signal intensity over a number of discrete frequency
channels, rather than integrating over the whole bandwidth before data recording, is so
that data contaminated by radio frequency interference (RFI) can be identified
and eliminated from any integration. Most RFI is prevalent in particular
channels which would dominate the integration were they not removed first.


\section{The Dynamic Spectrum Pipeline and Radio Frequency Interference (RFI)
Mitigation}
\label{sec:rfi}

A dynamic spectrum software pipeline is currently under construction: It needs
to process the data in a number of steps to correct the spectra for the antenna
response across the pass-band, identify channels mostly contaminated by RFI and
more random bad data points, and remove these from the resulting dynamic
spectra. These `corrected' data are then integrated across the pass-band to
create the IPS time series.

\begin{figure}
	\centering
	\includegraphics[width=10cm]{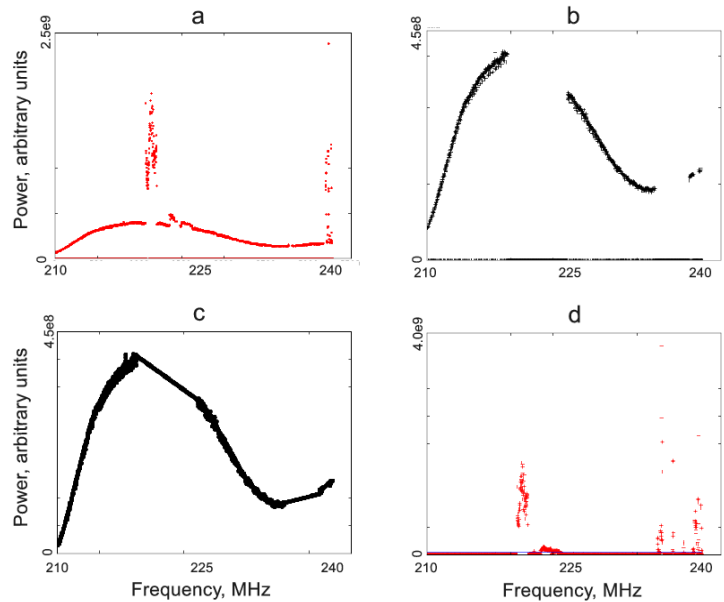}
\caption{RFI excision steps.  The frequency range 210--240\,MHz is used in this example.  The plots are of the median of the time series of
each frequency channel: (a) Constructed bias for channels; (b) We compute the
running median and local standard deviation, and use thresholding to identify
bad channels; (c) The bad channels are omitted and a smooth curve is fitted,
with linear interpolation, to obtain the spectral response of the system; (d)
The spectral response can now be subtracted from the input data.
The thick arrow displays the final threshold computed for the data.}
	\label{fig:rfi_excision}
\end{figure}

We look at the LOFAR data output as a time-frequency matrix consisting of M time
steps and N channels. Such a matrix is read in blocks of a length chosen by the user (20\,s has been chosen in the results presented here, the data arrays become unwieldy if a much higher length is chosen). The first
block is then processed to calculate the spectral reponse across the pass-band (illustrated in Figure~\ref{fig:rfi_excision}) and remove interference using the steps as follows:
\begin{enumerate}
\item We compute median spectrum from the time-frequency matrix.
\item For each channel we compute a `bias'  using the kurtosis of the time
samples in that channel, the standard deviation and the difference of the
median from medians of nearby channels. We apply a moderate clipping 
on this bias to  provide us with the first guess of channels heavily contaminated
by RFI (Figure~\ref{fig:rfi_excision}a).
\item A 1-d array of the medians of the time series' for each
channel is computed and further RFI-dominated channels identified using a 1-d walk through this array. Typically a small 32-channel window is used for this step to identify interfering sources which affect only a couple of neighbouring channels at most. Any channel with an anomalously large deviation from the median of the window is flagged.
\item The previous step is repeated with a large frequency window ($\sim 256$ channels), to
identify wide-band interfering sources. At this point most, if not all, of the
remaining RFI-affected channels are identified (Figure~\ref{fig:rfi_excision}b).
\item The RFI-laden channels are then omitted from processing. The remaining
channels are fitted with a smooth function in the frequency dimension after simple linear interpolation, which can then be used to remove the spectral response from the input time-frequency array (Figure~\ref{fig:rfi_excision}c).
\item A simple threshold of significance (usually $5\sigma$) above the (local or
global) median value of the flattened time-frequency array is then applied to
identify remaining RFI (Figure~\ref{fig:rfi_excision}d).
\item The time series for IPS is calculated from the time-frequency matrix after
zeroing the identified RFI-contaminated channels and other RFI points.
\end{enumerate}

In subsequent blocks the data are flattened using the same spectral response curve calculated for the first block and then processed according to the final two steps given above.

An option to calculate a `clean' 2D data matrix without `zeros' is also provided. To achieve this every `bad pixel' in the matrix is replaced by a random `good' pixel from the surrounding ($8 \times 32$) pixel region. Whilst the good pixels are not modified by this method, it undoubtedly introduces more noise to the
array and would not be safe to use when calculating the IPS time series.
However, it does provide a useful way by which general trends across the array
can be observed in a more convenient fashion. It will also allow 2D analysis of
the complete observed time-frequency data without any obvious artefacts
produced by the zero-mask. Since LOFAR offers a large fractional bandwidth,
the 2D dynamic spectra contain more information than mere time series.
The random substitution mode therefore may be relevant for the analysis of 2D
spectra.

\begin{figure}
	\centering
	\includegraphics[width=10cm]{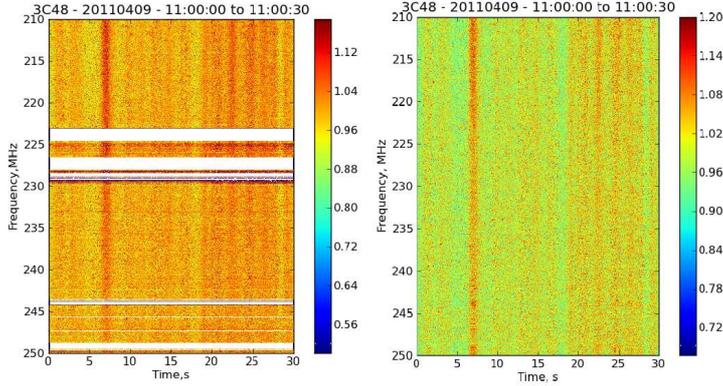}
\caption{The dynamic spectrum of the first 30s of data from an observation
of 3C48 taken on 9th April 2011. Left: The identified RFI points are left blank.
Right: The RFI pixels are replaced by random substitution from the
neighbourhood.}
	\label{fig:spectra_random_substitution}
\end{figure}

Figure~\ref{fig:spectra_random_substitution} displays the output data processing
for an observation of 3C48 taken in April 2011. The pixels wihin the RFI-affected 
regions are either blanked out or substituted with a random good neighbouring pixel.

\section{Results}
\label{sec:results}

\begin{figure}
	\centering
	\includegraphics[width=10cm]{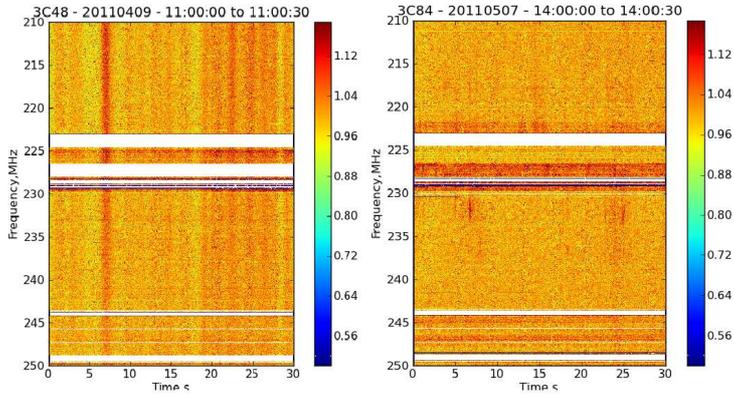}
\caption{Left: The dynamic spectrum of the first 30\,s of data from an observation
of 3C48 taken on 9th April 2011. Right: The same for an observation of 3C84
taken on 7th May 2011. Identified RFI is blanked out.}
	\label{fig:dynspec}
\end{figure}

Whilst the dynamic spectrum pipeline is still under development, initial analyses of
LOFAR commissioning observations of IPS are encouraging. Two examples of 30
seconds worth of data are given in Figure \ref{fig:dynspec}.

Clear differences can be seen in the dynamic spectra from these two
observations: In the observation of 3C48, clear bands of maxima and minima
corresponding to the IPS signal are seen right across the pass-band, though it
can also be noticed that the levels of the maxima appear to diminish at higher
frequencies. The IPS signal is not so readily apparent in the dynamic spectrum of the 3C84 observation; the scintillation only becomes apparent in the 1D time series and further observations since suggest that this is more the norm.

\begin{figure}
	\centering
	\includegraphics[width=10cm]{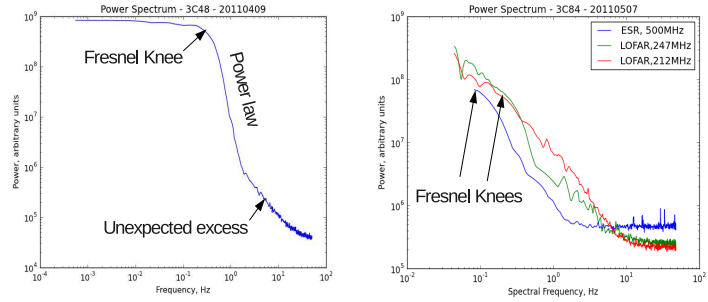}
\caption{Left: The power spectrum of the time series integrated over the
band-pass for the LOFAR observation of 3C48 taken on 9 April 2011. Right: The power spectra of time series' calculated for the highest and lowest 5\,MHz of the total LOFAR frequency band for the LOFAR observation of 3C84 taken on 7 May 2011, alongside the power spectrum for a simultaneous observation taken on the ESR.  The low-frequency parts of these spectra (usually ignored in analysis) have been removed for ease of viewing. }
	\label{fig:spectra}
\end{figure}

In both example observations, power spectra from the resulting time series have been
produced and are shown in Figure \ref{fig:spectra}. A simultaneous observation
of 3C84 was also taken using the EISCAT (\textit{European Incoherent SCATter}) \textit{Svalbard Radar} (ESR) in the high Arctic at a higher observing frequency of 500\,MHz and this is also shown for comparison; in this case two LOFAR spectra are shown, each corresponding to 5.4\,MHz of bandwidth from the upper and lower frequencies of the observation to match the bandwidth of the ESR observation.

It is expected (P.K. Manoharan, private communication, 2011) that the Fresnel knee for a lower-frequency observation will be at a slightly lower spectral frequency, but the power laws of each spectrum are expected to be similar.  Neither of these expectations appears to be met for the 3C84 observations (Figure \ref{fig:spectra}), although the power laws are arguably similar for the higher-frequency LOFAR and ESR power spectra.  The unexpected excess of power at higher frequencies apparent in the 3C48 observation is only apparent in the higher-frequency LOFAR power spectrum of 3C84.  Unfortunately, further comparisons on different observations have not proved possible to date.

\begin{figure}
  \centering
  \includegraphics[width=10cm]{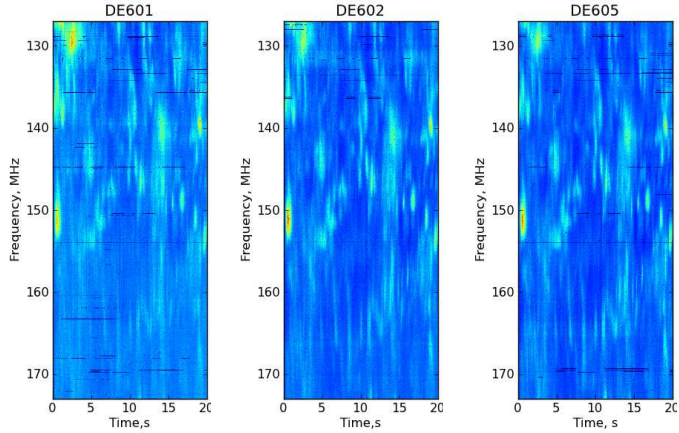}
  \caption{Dynamic spectra of the first 20\,s of data taken on three LOFAR stations as part of an observation of 3C279 on 1 October 2011.}
  \label{fig:20111001}
\end{figure}

Figure \ref{fig:20111001} shows the dynamic spectrum for three LOFAR stations of the first 20 seconds of data of an observation of 3C279 taken on 1st October 2011.  This observation was taken using the lower end of the HBA frequency range, centred on 150\,MHz at a time when 3C279 was only 8$^{\circ}$ away from the Sun.  This is well into the `strong'-scattering regime (a regime where it can no longer be assumed that the scattered radio waves do not interfere amongst themselves) for IPS at this observing frequency and the dynamic spectrum certainly appears to show structure which are consistent with that.  The larger structures at lower observing frequencies may be indicative of refractive (as opposed to diffractive) scintillation from large-scale density variations in the solar wind.  However, the dynamic spectra from the three stations all show exactly the same structures at exactly the same times.  

Further observations taken that day and in July also showed a high degree of correlation with zero time delay in the data between different LOFAR stations.  The cause of this is still being investigated.

\begin{figure}
  \centering
  \includegraphics[width=10cm]{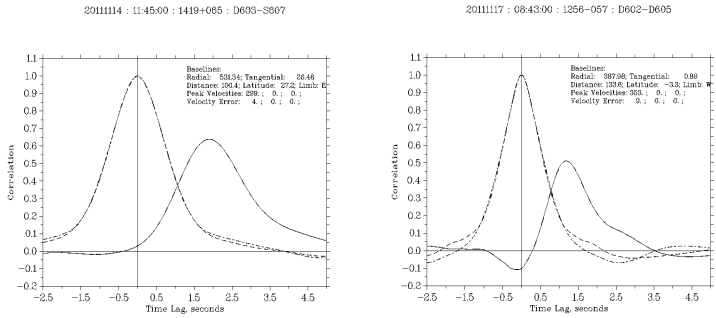}
  \caption{Left: Auto- and cross-correlation functions for an observation of 3C298 taken on 14 November 2011.  Right: Same for an observation of 3C279 taken on 17 November 2011.}
  \label{fig:correlations}
\end{figure}

Observations taken with multiple stations on 14 and 17 November 2011 show cross-correlation functions with time delays corresponding to those expected for a slow solar wind stream.  Two correlation functions are shown in Figure \ref{fig:correlations}, from observations of 3C298 on 14th November and of 3C279 on 17th November 2011.

The observation on 14 November indicates a slow solar wind stream travelling at approximately 300\,km\,s$^{-1}$; the negative lobe near zero time lag in the cross-correlation function of the 17th November observation may indicate the presence of a Coronal Mass Ejection (CME) in the lines of sight.


\section{Conclusions}
\label{sec:conclusions}

The dynamic spectrum results obtained so far hint at a wealth of new information
on solar wind micro-structure and turbulence. Previous studies are few, but
\citet{ColeSlee:1980} did observe a dynamic spectrum over the frequency range
280--520\,MHz of IPS seen in an observation of 3C273. This study showed a curve
in the scintillation maxima over the frequency band, later attributed to
refraction due to large-scale components of a Kolmogorov turbulence regime in
the solar wind \citep{ColesFilice:1984}. Such a curve is not seen in LOFAR
observations of IPS taken so far, most likely because the 40--48\,MHz bandwidth
used is not large enough for it to be seen clearly.

The level of scintillation is known to vary with both distance from the Sun and
observing frequency \citep[e.g.][]{Coles:1978}; within the `weak' scattering
regime, lower observing frequencies will exhibit stronger scintillation at the
same solar elongation. The observation of 3C48 shown in Figure \ref{fig:dynspec}
illustrates this nicely, with the scintillation being stronger at the lower
frequencies.

The power spectra shown in Figure \ref{fig:spectra} are consistent with typical
IPS spectra, but also show some inconsistencies. The excess power evident at
higher spectral frequencies in the observation of 3C48 and the inconsistency in
the observation of 3C84 when compared to a simultaneous observation from a known
antenna are particular points of concern. It is possible that at least some of
these inconsistencies are due to the high LOFAR observing frequencies used: It is known that a grating lobe of the main station beam (an artifact of the use of an array of dipoles) can be present above the horizon for frequencies in the high-band of LOFAR (Ger de Bruyn, private communication, 2011), potentially causing issues with excess noise.  This is not well-characterised yet making predictions of whether it will or will not be above the horizon for a particular observation difficult.  It is clear from all the observations of IPS taken so far that the excess power noted in the observation of 3C48 in Figure \ref{fig:spectra} is apparent in many observations using the HBA, but not in all.  Restricting the sub-bands used in creating the time series to those at the lower or higher ends of the observing band may occasionally make a difference but, again, not in every case.  

The correlation functions shown in Figure \ref{fig:correlations} indicate the presence of slow solar wind streams in the lines of sight.  This is consistent with the low heliographic latitudes of both these observations.  The cross-correlation function of the 3C279 observation shows a `negative lobe' at zero time-lag.  This is often associated with the presence of a Coronal Mass Ejection in the line of sight.  A slow CME was observed to launch from the Sun late on 14 November 2011 and was predicted to pass close to the Earth on 18-19 November 2011.  An initial check of coronagraph data indicate that this CME was launched in a direction that could cross the line of sight of the IPS observation and the observation timing makes it a promising candidate.  However a full geometrical analysis will be required to confirm whether or not this CME is seen in these IPS data.

In conclusion, these observations show a high degree of promise, but also reveal that some issues remain.  This is to be expected from an instrument which is still undergoing commissioning.


\begin{acks}
LOFAR, the \textit{Low Frequency Array} designed and constructed by ASTRON, has facilities in several countries, that are owned by various parties (each with their own funding sources), and that are collectively operated by the International LOFAR Telescope (ILT) foundation under a joint scientific policy.  The authors thank the director and staff of EISCAT for the ESR data used in this study.  EISCAT is funded by the research councils of Norway, Sweden, Finland, Japan, China, the United Kingdom and Germany.  Two of us (RAF and MMB) were funded by the UK Science and Technology Facilities Council during the course of this work.
\end{acks}


\end{article}

\end{document}